\documentclass[twocolumn]{aastex631}

\usepackage{threeparttable}
\usepackage{parskip}
\usepackage{float}
\usepackage{hyperref}
\usepackage{graphicx}

\usepackage{amsmath}

\usepackage{soul}

\newcommand{\nicer}{{\it NICER\/}} %

\newcommand{\xmm}{{\it XMM-Newton\/}}
\newcommand{\swift}{{\it Swift\/}}
\newcommand{\maxi}{{\it MAXI\/}} %

\begin{document}

\title{Discovery of a Long Thermonuclear X-ray Burst from the Ultracompact Binary 4U~1850--087}

\author{Yongqi Lu}
\affiliation{Key Laboratory of Stars and Interstellar Medium, Xiangtan University, Xiangtan 411105, Hunan, P.R. China}

\correspondingauthor{Zhaosheng Li, Yuanyue Pan}
\email{lizhaosheng@xtu.edu.cn, panyy@xtu.edu.cn}

\author[0000-0003-2310-8105]{Zhaosheng Li}
\affiliation{Key Laboratory of Stars and Interstellar Medium, Xiangtan University, Xiangtan 411105, Hunan, P.R. China}

\author{Wenhui Yu}
\affiliation{Key Laboratory of Stars and Interstellar Medium, Xiangtan University, Xiangtan 411105, Hunan, P.R. China}

\author{Yuanyue Pan}
\affiliation{Key Laboratory of Stars and Interstellar Medium, Xiangtan University, Xiangtan 411105, Hunan, P.R. China}

\author[0000-0003-3095-6065]{Maurizio Falanga}
\affiliation{International Space Science Institute (ISSI), Hallerstrasse 6, 3012 Bern, Switzerland
}
\affiliation{Physikalisches Institut, University of Bern, Sidlerstrasse 5, 3012 Bern, Switzerland
}

\begin{abstract}
We report the detection of a long X-ray burst from the ultracompact binary 4U~1850--087. The type I X-ray burst was observed on MJD 60171.65 by the Monitor of All-sky X-ray image and Neutron Star Interior Composition Explorer NICER. We analyze the NICER data between MJD 60095.19 and 60177.43, which includes the observation covering parts of the long X-ray burst decay phase, i.e., 0.15--3.8 hr after the trigger. The persistent spectra are well described by a multicolor disk blackbody, with an inner temperature of $\sim 0.5$ keV, a thermally Comptonized continuum, an asymptotic power-law photon index of $\Gamma\sim2.2$, and an electron temperature of 20-30 keV. The mean persistent flux was around $3.8\times10^{-10}~{\rm erg~cm^{-2}~s^{-1}}$, corresponding to a local mass accretion rate of $\sim 1\%~\dot{m}_{\rm Edd}$. Part of the time-resolved burst spectra show a clear deviation from the blackbody model, which can be improved by considering the enhanced persistent emission due to the Poynting-Robertson drag  or the reflected disk emission illuminated by the burst. From the burst flux during the cooling phase, we estimate the burst duration, $\tau \approx 0.78$ hr, the burst fluence, $E_\mathrm{bb} \approx $4.1$ \times 10^{41}$ erg, and the ignition column depth, $y_{\rm ign}\approx$3.5$\times10^{10}~{\rm g~cm^{-2}}$. These  long X-ray burst parameters from 4U~1850--087 suggest a regime of unstable burning of a thick, pure helium layer slowly accreted from a helium donor star. Moreover,  we identify $7\sigma$ significant  $\sim1$ keV emission lines in the burst spectra, which may originate from the surrounding accretion disk. 
\end{abstract}

\keywords{}

\section{Introduction} \label{sec:intro}
In a low-mass X-ray binary  (LMXB) hosting a neutron star (NS), matter is accreted from a companion star ($M<1M_\odot$) through the Rochelobe overflow. NS LMXBs can be observed as a persistent or transient systems \citep{Lasota01}. The accreted matter, mainly composed of helium or mixed hydrogen and helium, can be consumed through thermonuclear stable, marginally stable, or unstable burning. These processes depend on the mass accretion rate and the abundance of the chemical elements \citep{keek12,Li21}. The unstable thermonuclear burning is also known as type I X-ray burst \citep[hereafter X-ray burst; see][]{Strohmayer06,Galloway21}{}{}. X-ray bursts usually release a total energy of $\sim 10^{39}$ erg and ignite at a typical column depth of  $y_{\rm ign}\sim 10^8 \mathrm{~g~cm^{-2}} $. They lasts $\sim$ 10--100 s,  depending on the thickness of the burning layer ~\citep[see ][]{Lewin93,galloway2008thermonuclear,Galloway20}{}{}. 
Intermediate-duration bursts can release $\sim 10^{41} \mathrm{~erg}$  in a longer duration of $\sim$ 100--1000 s and are powered by unstable burning of helium in \textbf{a} deep layer at an ignition column depth of $y_{\rm ign}\sim 10^{10} \mathrm{~g~cm^{-2}}$ \citep[see ][]{Zand05,Cumming06,Falanga08,Keek10}{}{}. Most intermediate-duration bursts occurred in the low mass accretion rate, which is lower than 1\% of the Eddington limit \citep{Falanga08,Alizai23}. 
Superbursts have the longest duration of $>10^3$  s and are powered by thermonuclear unstable burning of carbon, with a mass carbon mass fraction reaching  15\%--30\%, and releasing energies of $\sim 10^{42} \mathrm{~erg}$ ~\citep{Cumming01,Strohmayer02}. However, some of the X-ray bursts in 4U~0614+091 and SAX J1712.6--3739 with a duration of hours were classified as intermediate-duration bursts since the mass accretion rates of  $\sim1$\% of the Eddington limit cannot produce sufficient carbon to trigger a superburst \citep[see,e.g.,][]{Kuulkers10,zand19A&A,Alizai23}.

Several types of spectral features have been identified during  X-ray bursts. The 1 keV emission lines have been observed in IGR J17062--6143 \citep{Degenaar_2013, keek17, Bult21} and SAX J1808.4--3658 \citep{2021ApJ_Bult}, which were interpreted as a reflection feature from the accretion disk, possibly associated with Fe L-band transitions. The 1 keV emission line was also observed from 4U 1820--30 during a burst expansion phase  \citep{Strohmayer19}. In addition, disk reflection features during an X-ray burst were also observed without the 1 keV emission in several normal or long X-ray bursts \citep{Strohmayer02,Ballantyne04,keek14,Zhao22,lyq23A&A}. 

4U~1850--087 was discovered as a persistent X-ray source and classified as a X-ray burster hosting a NS \citep{swank76}. 
The possible 20.6-minute orbital period associated 4U~1850--087 whit an ultracompact X-ray binary \citep[UCXB;][]{homer96}. 4U 1850--087 is located in the globular cluster NGC 6712 at a distance of 8 kpc \citep{18Tremou}. 
The  \xmm\ persistent spectra detected O K, Fe L, and Ne K edges at 0.54, 0.71, and 0.87 keV, respectively \citep{Sidoli05}. 
Several long X-ray bursts from 4U~1850--087 were detected by the Monitor of All-sky X-ray Image (MAXI) and \swift.  These bursts were triggered by unstable pure helium burning at low mass accretion rate and in a deep layer,due to the high total energy released \citep[e.g.,][]{zand14,Serino16}{}. 

In this work, we analyze a long X-ray burst detected by nicer and MAXI on 2023 August 15. In Section \ref{sec:ob}, we introduce the observations and describe the persistent emissions and long X-ray burst properties. In Section \ref{sec:Ana}, we perform the spectral analysis of the persistent and time-resolved burst emissions. We discuss the results in Section \ref{sec:Dis} and summarize the findings in Section~\ref{sec:sum}.

\section{Observation and Data Reduction} \label{sec:ob}

MAXI and NICER are two instruments installed on board the International Space Station that have been operating since 2009 July 15 and 2017 June 3, respectively. To monitor the X-ray  activity, such as X-ray outbursts and X-ray bursts, MAXI performs a full-sky survey every 96 minutes. From the MAXI novae webpage\footnote{\url{http://maxi.riken.jp/alert/novae/}}, we noticed a long X-ray burst from 4U~1850--087 triggered on August 15, 2023 (MJD 60171.654514). We then cross-checked the NICER archived data and found six contemporaneous observations, including Obs. IDs 6617010101--6617010103 and 6403520101--6403520103, with a net unfiltered exposure time of 44 ks. The Obs. ID 6403520101 started on MJD 60171.66085, i.e., 0.15 hrs after the trigger of MAXI, partially covering the cooling tail of the long X-ray burst.

\begin{figure}[ht!]
\includegraphics[width=\linewidth]{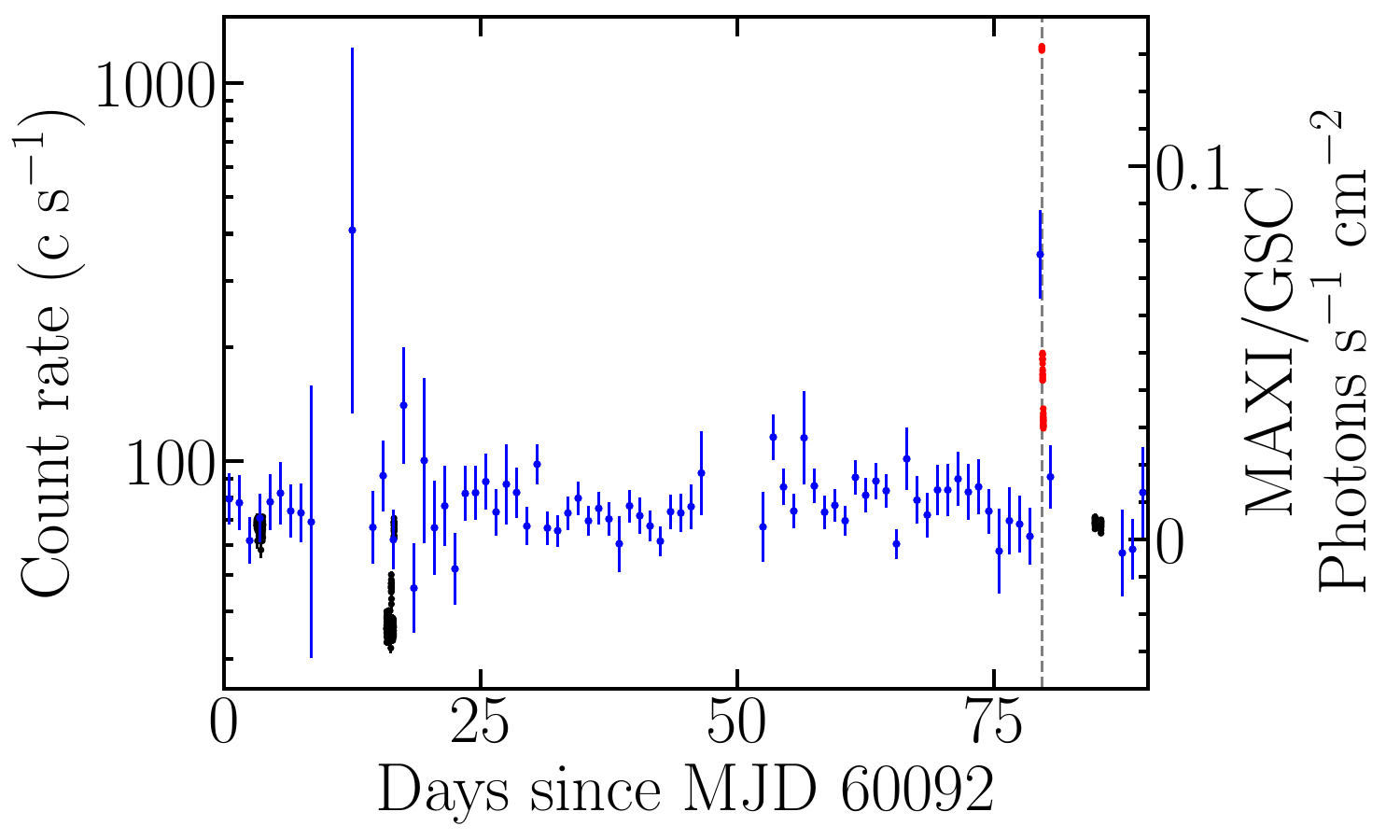}
\caption{ Light curves of 4U~1850--087 from MAXI and NICER observations. We show the  \nicer\ light curve (64 s, 0.5--10~keV; black and red points) and the MAXI light curve (one day averaged, 2--20~keV; blue points), where the red points represent the NICER long X-ray burst's data.
The vertical gray dashed line on MJD 60171.654514 represents the MAXI burst trigger time.}
\label{fig:lc}
\end{figure}

We used HEASOFT V6.31.1 and the NICER Data Analysis Software (NICERDAS) to process the NICER data. The standard filtering criteria have been applied, i.e., the pointing offset $<54''$, the Earth elevation angle  $>15^\circ$, the elevation angle with respect to the bright Earth limb $\leq30^\circ$,  and the instrument \textbf{located} outside the South Atlantic Anomaly (SAA). 
We used \texttt{nicerl3-lc} to extract the light curves in 0.5--10 keV, 0.5--2keV, and 2--10 keV. The 64 s light curves in the energy range 0.5--10 keV are shown in Figure~\ref{fig:lc}. 
To have a better coverage of the data between MJD 60171 and 60172, we added the 2--20 keV and 1-day binned light curves from \maxi/GSC\footnote{\url{http://maxi.riken.jp/mxondem/}} (see Figure~\ref{fig:lc}).  

\begin{figure}[ht!]
\includegraphics[width=\linewidth]{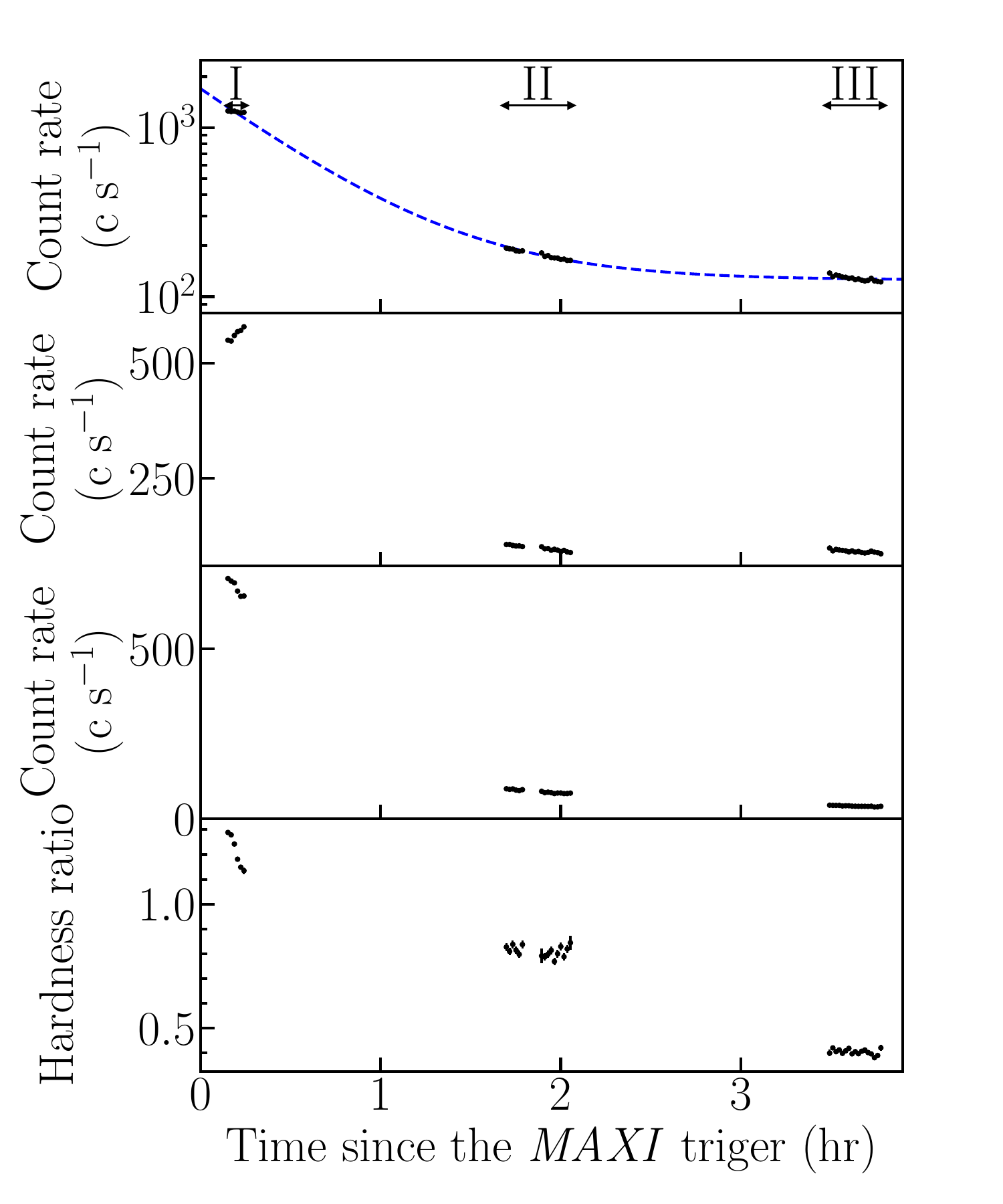}
\caption{The burst light curves of 4U~1850--087 from NICER ObsID 6403520101. From top to bottom, we show the 64 s NICER burst light curves in the energy ranges 0.5--10~keV, 0.5--2~keV, and 2--10~keV and the hardness ratio between 2.0--10 and 0.5--2 keV, respectively. The top panel shows} the fitted exponential model with the blue dashed line. 
The time intervals of regions I, II, and III are represented. 
\label{fig:lc_burst}
\end{figure}

We show the 64 s light curves of 0.5--2keV, 2--10keV, and 0.5--10 keV from ObsID 6403520101 in Figure~~\ref{fig:lc_burst}. The hardness ratio between 2--10 and 0.5--2 keV is displayed in the bottom panel in Figure~\ref{fig:lc_burst}. The light curve contains three nearly continuous segments, i.e., 0.15--0.25, 1.7--2.05, and 3.5--3.8 hr after the MAXI trigger,  marked as regions I, II, and III, respectively. The count rate of ObsID 6403520101 decreased from 1250 to 120 counts s$^{-1}$ in 3.6 hr.  
The count rate in region III is higher than the persistent count rate of about 70 counts s$^{-1}$, therefore, the three regions are all considered part of the burst.
Meanwhile, the hardness ratio decreased from 1.3 to 1.1 in region  I and was unchanged at a level of 0.8 and 0.4 in regions II and III, respectively, which indicates the spectral softening during the burst cooling; see Sect.~\ref{subsec:burst}. We model the 1 s burst decay light curve in the energy range 0.5--10~keV by an exponential function 
\begin{equation}
C(t)=C(t_0)e^{-(t-t_0)/\tau_{\rm LC}}+C_0. 
\label{eq:exp}
\end{equation} 
The model includes the normalization, $C(t_0)$, the exponential decay time, $\tau_{\rm LC}$, and a constant, $C_0$, representing the burst's persistent count rate. We fix the $t_0$ to the MAXI trigger time. The fitted model gives an exponential decay time $\tau_{\rm LC}=0.55\pm0.01$ hr, with a $\chi_{\nu}^2{(\rm dof)}=1.35(2306)$. We show the best-fitted model in the top panel of Figure~\ref{fig:lc_burst}.

We applied the fast Fourier transform method with Leahy normalization \citep{Leahy83} to search for burst oscillation in ObsID 6403520101. The power spectra were calculated in the 0.5--3keV, 3--10 keV and 0.5-10keV energy ranges from the cleaned NICER event file. We adopted a moving window method with the window size $T$ of 8 s, with steps of 1 s. We calculated the Fourier frequency between 50 and 2000 Hz with an interval of 0.25 Hz in each window \citep[see, e.g.][]{Bilous19,Li21}{}. We found no  burst oscillation signal in NICER observations and put an upper limit on the fractional rms amplitude of $\sim 5\%$.

\section{Spectral Analysis} 
\label{sec:Ana}

We extracted the NICER spectra, 3C50 background spectra,  ancillary response files (ARFs), and response matrix files (RMFs) using \texttt{nicerl3-spect}~\citep{Remillard21}. 
We used XSPEC v12.13.0 to analyze the NICER spectra in the 0.7--10 keV energy band~\citep{Arnaud96}. 
All bolometric fluxes were estimated in the energy range 0.01-250 keV by using the \texttt{cflux} model. The uncertainties are reported at $1\sigma$ confidence level.

\subsection{Persistent Emissions} \label{subsec:presb}

We fitted the persistent spectra from ObsID 6403520102--6403520103 and ObsID 6617010101--6617010103 individually. All persistent spectra were grouped with at least 50 counts per channel. The persistent spectra can be well described by a combination of a disk blackbody plus a high-energy Comptonized component \citep[see,e.g.][]{Sidoli06}{}. We adopted the model  \texttt{TBabs*(diskbb+nthcomp)} to fit the spectra, where  \texttt{TBabs} is the interstellar absorption  with abundances from \citet{Wilms_2000}, \texttt{diskbb} is a multidisk blackbody spectrum; \texttt{nthcomp} is a thermally Comptonized continuum \citep{Zdziarski96,Zycki99}. 
For \texttt{nthcomp}, the parameters are the asymptotic power-law index, $\Gamma$, the hot electron temperature, $kT_\mathrm{ee}$, the blackbody (multicolor disk) seed photon temperature, $kT_\mathrm{bb,~seed}$, the input type of seed photons, and the normalization. For \texttt{diskbb}, the parameters are the inner disk temperature,  $kT_\mathrm{in}$ and the normalization. 
For \texttt{TBabs} the free parameter is the equivalent hydrogen column, $N_{\rm H}$.  We assume the input seed photons of \texttt{nthcomp} coming from the accretion disk and tied the $kT_\mathrm{bb,~seed}$ to $kT_\mathrm{in}$ of the \texttt{diskbb} model. Considering the long exposure time for a few spectra, the model fitted the spectra well with the reduced $\chi^2$, $\chi^2_{\nu}\sim0.9-1.3$. 
The best-fitted parameters are listed in Table~\ref{table:preburst}. 

\begin{figure}[ht!]
\includegraphics[width=\hsize]{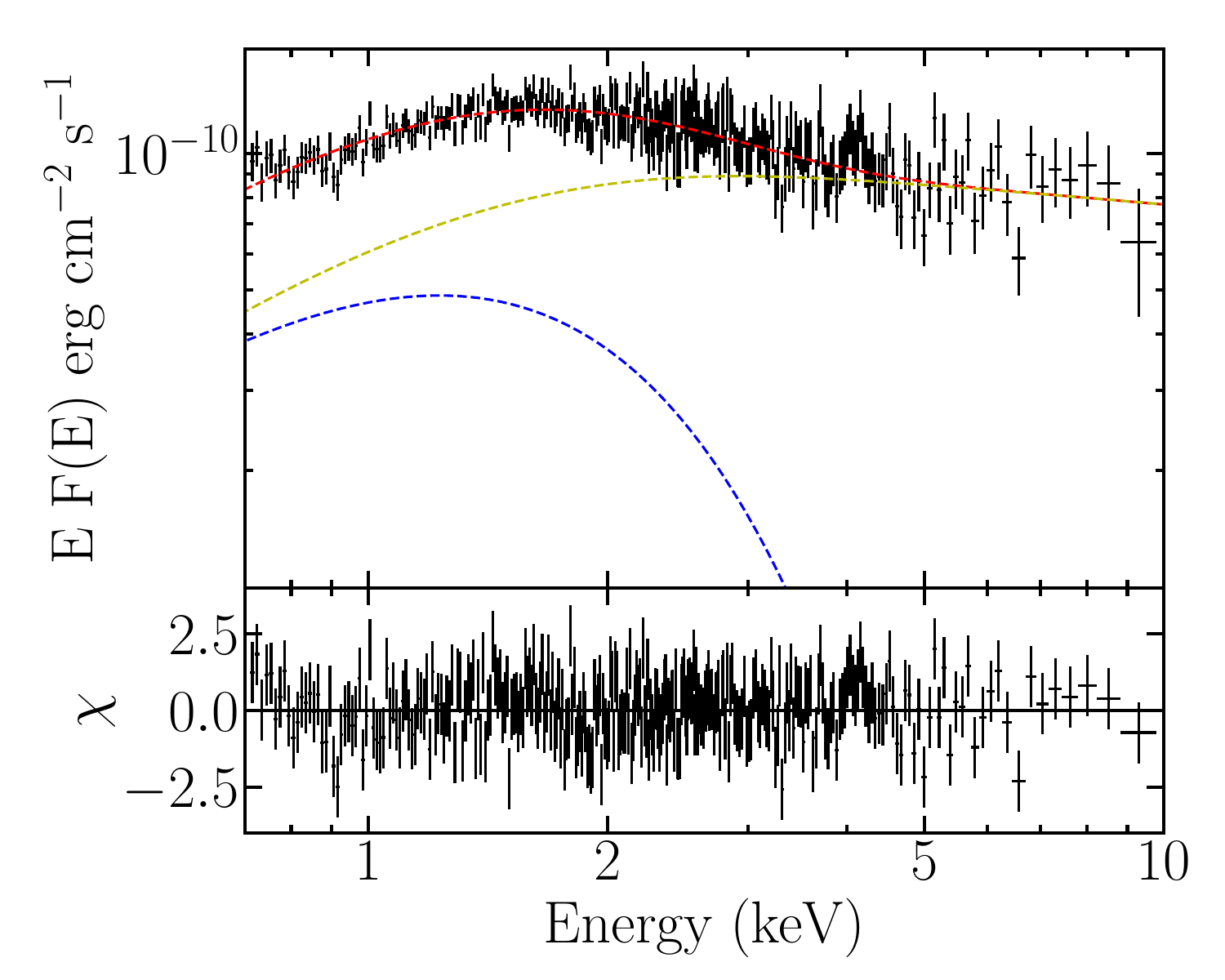}
\caption{The un-absorbed persistent spectrum  and residuals from \nicer\ Obs. ID 6403520102 in 0.7--10 keV was regarded as the persistent emission during the long X-ray burst. The red, yellow, and blue dashed dots present the model, $\texttt{nthcomp}$ component, and $\texttt{diskbb}$ component, respectively. }\label{fig:preburst}
\end{figure}

 The averaged $N_{\rm H}=0.61 \pm{0.01} \times 10^{22}~{\rm cm} ^{-2}$ is consistent within the range of 0.4--0.6 $\times 10^{22}~{\rm cm} ^{-2}$ from \citet{Sidoli06}. 
The averaged disk temperature, radius, and power-law index of the \texttt{nthcomp} component are  $kT_{\rm in}= 0.48\pm0.07$ keV, $R_{\rm in}=7\pm1$ km, and $\Gamma= 2.2\pm0.1$, respectively. The plasma electron temperature is hard to constrain tightly, and the uncertainties are large.
These five independently fitted persistent spectra showed quite similar parameters, implying that the persistent emissions did not change significantly. Therefore, we adopted the spectrum from Obs. ID 6403520102, which was close to the long X-ray burst, accounts for the persistent emission during the burst. We show the best-fitted spectrum and residuals in Figure~\ref{fig:preburst}.

\begin{table*}
\begin{center} 
\caption{ Best-fitted parameters of All persistent spectra.  \label{table:preburst}}
\resizebox{\linewidth}{!}{\begin{tabular}{ccccccccc} 
\hline\\ %
{\centering NICER} &
\centering  Exposure&
{\centering  $N_{\rm H}$  } &
{\centering  $\Gamma$} &
{\centering  $kT_{\rm \textit{e}}^{\mathrm{a}}$} &
{\centering  $kT_{\rm in}$} &
{\centering  $R_{\rm in}^{\mathrm{b}}$} &
{\centering  $F_{\rm per}^{\mathrm{c}}$} &
{\centering  $\chi_{\nu}^{2}(\mathrm{dof}$)}
\\
(ObsId)&(s)&$(10^{21}\enspace  \mathrm{cm^{-2}})$ & & $\mathrm{(keV)}$ & $\mathrm{(keV)}$ &(km)& ($10^{-10}~{\rm erg~cm^{-2}~s^{-1}}$) &\\ [0.01cm] \hline
6403520102&647    &5.92$\pm{0.17}$& $2.13^{+0.10}_{-0.18}$&$18.46_{-17.30}^{+p}$& $0.53\pm{0.04}$&$6.76^{+1.32}_{-0.63}$&$3.86\pm{0.03}$& 0.89(294)\\
6403520103&1533   &5.72$\pm{0.12}$& $2.07^{+0.08}_{-0.09}$&$28.50_{-27.50}^{+p}$& $0.56\pm{0.03}$&$6.08^{+0.38}_{-0.42}$&$3.79\pm{0.02}$& 1.05(403)\\
6617010101&8833   &5.98$\pm{0.11}$& $2.13^{+0.04}_{-0.04}$&$31.31_{-30.05}^{+p}$& $0.49\pm{0.02}$&$7.57^{+0.27}_{-0.29}$&$3.82\pm{0.02}$& 1.29(643)\\ 
6617010102&4492   &6.32$\pm{0.09}$& $2.38^{+0.03}_{-0.08}$&$18.31_{-17.25}^{+p}$& $0.41\pm{0.03}$&$7.14^{+1.03}_{-1.79}$&$3.75\pm{0.02}$& 1.24(454)\\ 
6617010103&3931&6.40$\pm{0.12}$& $2.27^{+0.06}_{-0.08}$&$14.70_{-10.67}^{+p}$& $0.40\pm{0.04}$&$8.71^{+1.21}_{-1.51}$&$3.85\pm{0.02}$& 1.13(438)\\ 
\hline
\end{tabular} }
\end{center}
$^{\mathrm{a}}$ The symbol $p$ means that the poorly constrained electron temperature is pegged at the hard limit. %

$^{\mathrm{b}}$ We calculated the inner disk radius by assuming the angle of the disk, $i=30^{\circ}$.

$^{\mathrm{c}}$ The unabsorbed bolometric persistent flux in 0.1-250 keV.
\end{table*}  

\subsection{Time-resolved burst Spectroscopy\label{subsec:burst}}
We extracted the time-resolved burst spectra with time bins of 10 s in region I (total exposure time of 300 s and 100 s in regions II and III from the cleaned event file in ObsID 6403520101. 
We grouped the burst spectra to guarantee that each channel has at least 20 counts by using \texttt{grappha}. 
We first adopted the standard model, \texttt{TBabs $\times$ bbodyrad}, to fit the burst spectra, where the parameters are the temperature, $kT_{\rm bb}$, and normalization, $K=R_{\rm bb}/D_{\rm 10~kpc}$ where $D_{\rm 10~ kpc}$ is the distance to the source in units of 10 kpc, for \texttt{bbodyrad}, and $N_{\rm H}$ for \texttt{Tbabs}. We fixed $N_{\rm H}=0.6\times 10^{22}~{\rm cm} ^{-2}$ and took the persistent spectra from ObsID 6403520102 as the burst background. 
The blackbody model can well describe the burst spectra in regions II and III for $\chi_{\nu}^{2}$ close to unity. However, we obtained $\chi_{\nu}^{2}\sim1.8-2.3$ for region I, implying poor fits.
We show the spectral parameters and $\chi_{\nu}^{2}$ as grey stars in Figure~\ref{fig:burst}. The improvements of the fit can be obtained by introducing an extra component from the enhancement of persistent emission during the burst, i.e., the $f_a$-model (see Sect.~\ref{subsub:fa}), or a reflected emission from the accretion disk, i.e., the reflection model (see Sect.~\ref{subsub:re}).

\subsubsection{Enhanced Persistent Emission}
\label{subsub:fa}
\begin{figure}[ht!]
\includegraphics[width=\hsize]{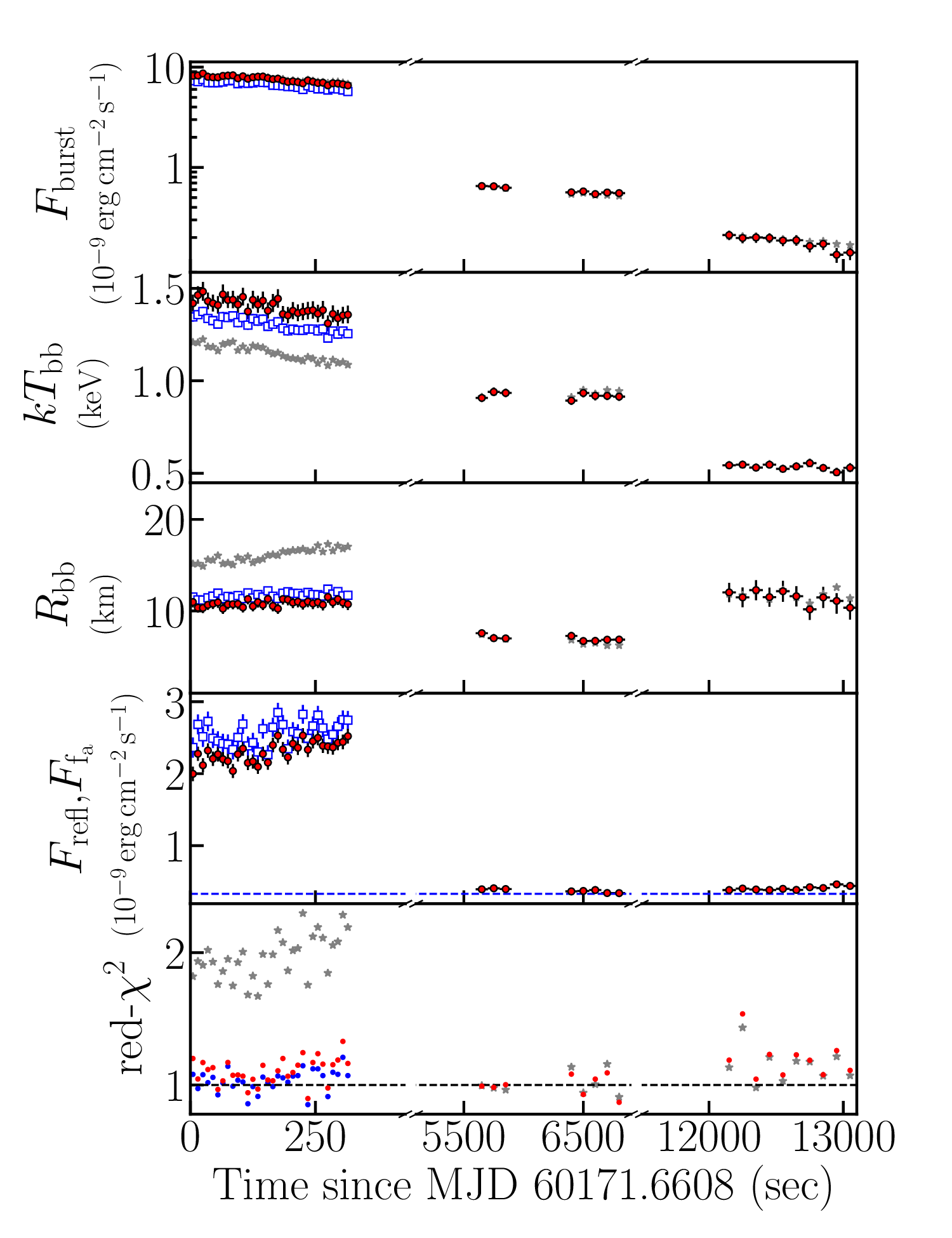}
\caption{The time-resolved spectroscopy of the long X-ray burst observed by NICER. All data points belong to the part of the long X-ray burst.  From top to bottom, we show the bolometric flux of burst, $F_\mathrm{ burst}$, the blackbody temperature, $kT_\mathrm{bb}$, and the blackbody radius, $R_{\rm bb}$, which were calculated at a distance of 8 kpc; the enhanced persistent flux, $F_\mathrm{f_a}$, or reflection flux, $F_\mathrm{refl}$; and the goodness of fit per degree of freedom, $\chi_{\nu}^{2}$. The gray star, red circle, and blue square display the results from the blackbody model, the $f_a$ model, and the reflection model, respectively. The blue dashed line represents the persistent emission level. The time zero corresponds to about 0.15 hr after the long burst starts.
}\label{fig:burst}
\end{figure}

We used the $f_{\rm a}$-model, \texttt{TBabs$\times$(bbodyrad +  constant $\times$ (nthcomp + diskbb))}, to fit the burst spectra, where the \texttt{constant} accounts for the enhancement of the persistent emission due to the Poynting-Robertson drag \citep{Walker92,Zand13,Worpel13, Worpel15}. Here we assumed that only the amplitude of the persistent emission changed if the accretion rate increased during the burst. 
We adopted the 3C50 spectrum as the instrumental background \citep{Remillard22}. We fixed $N_{\rm H}$ to the averaged value of $0.6\times 10^{22}~{\rm cm} ^{-2}$ obtained in Sect.~\ref{subsec:presb}, and the \texttt{diskbb}, and \texttt{nthcomp} parameters to the best-fitted values from the persistent spectra. 
We show the new fitted parameter and $\chi_{\nu}^{2}$ as red points in Figure~\ref{fig:burst}.  
The $\chi_{\nu}^{2}$ of the best-fitted $f_{\rm a}$-model is around 1.0, which implies that the enhancement of the persistent emission can significantly improve the fitted results. 
The blackbody temperature decreased from $\sim$ 1.5 keV  (region I) to $\sim$ 0.55 keV (region III) as expected during the burst cooling. As the temperature decreases, the blackbody flux decays from $1\times 10^{-8}$ to $1.5\times10^{-10}~{\rm erg~cm^{-2}~s^{-1}}$. The blackbody radius is constant at $\sim10$ km \textbf{in} all regions, implying that the whole NS surface is cooling. The enhanced persistent emission during the burst region I was about five times the preburst emission, While the burst persistent emission in regions II and III is close to the preburst level. Due to the data gap of the first 0.15 hr of the burst, we could not verify that a photospheric radius expansion occurred. 

\subsubsection{The reflection model\label{subsub:re}}
In this subsection, we describe the absorbed burst blackbody emission obtained in region I by adding a reflection component. We use the \texttt{relxillNS} model  \citep{22Garcia} to describe the reflection emission \citep[see also][]{Zhao22,lyq23A&A,ywh23}{}{}. We fitted the burst spectra by \texttt{TBabs} $\times (\texttt{bbodyrad} +  \texttt{relxillNS})$.  We tied the input blackbody temperature, $kT_\mathrm{bb,~in}$, to the temperature of \texttt{bbodyrad}, and set the reflection fraction parameter to -1, which means that this model only represents the reflection emission, as the accretion disk is illuminated by \texttt{bbodyrad} photons. 
We fixed the indices of emissivity, $q_1=q_2=3$, the inner radius of the accretion disk, $R_\mathrm{in}=6R_{g}$, and the outer radius of the accretion disk, $R_\mathrm{out}=400R_{g}$, where the $R_{g}$ is the gravitational radius. 
Since the spin frequency of 4U 1850--087 is unknown, we assume the spin parameter, $a$, of the model \texttt{relxillNS} to be 0.1. We also tried two typical inclination angles, $30^{\circ}$ and $60^{\circ}$, to fit the spectra, and we found that a smaller inclination provides slightly better $\chi_\nu^2$. Therefore, we use $30^{\circ}$ as the inclination angle.
We let the ionization parameter, $\log\xi$, the iron abundance in the accretion disk, $A_\mathrm{Fe}$, and the accretion disk density, $\log n$, all free to vary. We found that the values of $\log \xi$, $A_\mathrm{Fe}$, and $\log n$ during the burst did not vary.  Therefore, we fixed these parameters to the mean values of  2.8, 6 in units of solar abundance, and 16 cm$^{-3}$, respectively. 
We only show the parameters of \texttt{relxillNS} in region I. 
The $\chi_\nu^2$ are close to 1 in region I, meaning that the reflection model can also explain the blackbody derivations. The best-fitting results are shown in Figure~\ref{fig:burst} as blue squares.  %
Compared with the $f_{\rm a}$-model, the reflection model provides 5\% lower blackbody temperatures,  5\% higher blackbody radii, and thus 15\% lower bolometric fluxes.  

\subsection{The 1 keV emission line\label{sec:line}} 
\begin{figure}[ht!]
\includegraphics[width=\hsize]{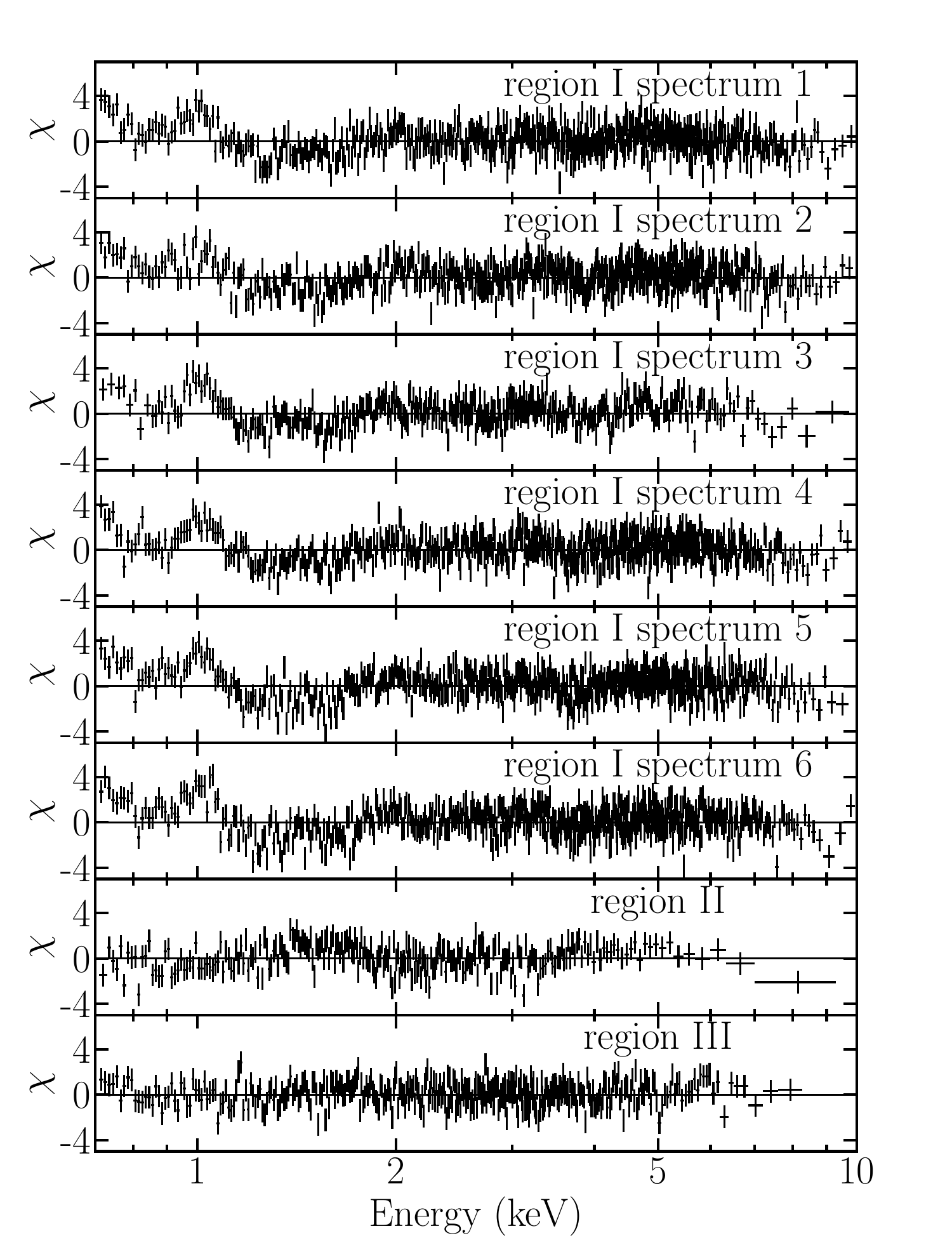}
\caption{The burst residuals obtained by fitting the $f_{\rm a}$-model.  We show error-weighted residuals from the regions I-III. The 1 keV emission features are clearly shown in the spectra from region I but not in the spectra from regions II and III.}
\label{fig:line}
\end{figure}

During the analysis of 30 time-resolved burst spectra, each of them lasting 10 s, we found a faint emission line around 1 keV in region I. In order to improve the significance of the emission line, we extracted six spectra in region I with an exposure time each of around 50 s. We used the $f_{\rm a}$-model to fit these spectra, as in section~\ref{subsub:fa}. We also extracted longer-duration spectra in regions II and III to find this possible 1 keV feature, but no significant features were found. We also found no prominent feature around 1 keV in the persistent spectra. We show the residuals of six spectra in region I and the residuals of spectra extracted from regions II and III as comparisons in Figure~\ref{fig:line}. 

We added a \texttt{Gaussian} component to account for the emission line. 
We used the \texttt{simftest} script in XSPEC to evaluate the statistical significance of the \texttt{Gaussian} component. The estimation contains $10^5$ sets of simulated data sets for each spectrum. We obtained a probability of the emission line arising by chance of less than $1.5\times10^{-14}$, which means that all lines' significance is higher than $7\sigma$. The Gaussian parameters of these six spectra are shown in Table~\ref{table:line}.  By using the reflection model, we also observed this 1 keV feature at similar significance.

\begin{table}
\begin{center} 
\caption{Gaussian line parameters. \label{table:line}}
\begin{tabular}{ccc} 
\hline
\hline 
Spectra in Region I&$E_{\rm center}$&Line Norm
\\
&(keV)&$(\mathrm{photons~cm^{-2}~s^{-1} })$\\
\hline

1& 0.97$\pm{0.02}$ & 0.03$\pm{0.01}$\\
2& 0.98$\pm{0.01}$ & 0.04$\pm{0.01}$\\
3& 0.99$\pm{0.01}$ & 0.03$\pm{0.01}$\\
4& 1.00$\pm{0.01}$ & 0.03$\pm{0.01}$\\
5& 0.98$\pm{0.01}$ &0.04$\pm{0.01}$\\
6& 0.98$\pm{0.01}$ &0.04$\pm{0.01}$\\
\hline
\end{tabular} 
\end{center}

\end{table}

\section{Discussion and conclusion} \label{sec:Dis}

\subsection{Burst and persistent spectra}
In this work, we report the long X-ray burst from 4U 1850--087 observed by NICER and MAXI in 2023. The persistent spectra can be well described by a combination of a multicolor disk blackbody with an inner temperature of -0.5 keV, and a thermally Comptonized continuum with an asymptotic power-law photon index of $\Gamma\sim2.2$, and electron temperature 20-30 keV. 
The persistent spectra are similar across all NICER observations; therefore, we adopted a spectrum after the long X-ray burst to represent the persistent emission during burst. We extracted the light curve and identified a long X-ray burst from the ObsID 6403520101. We divide the observation into regions I, II, and III to describe the burst evolution according to the light curve. The spectra from regions II and III were well described by a single blackbody model, but the region I spectra showed clear derivations from the blackbody shape. The results can be improved by adding the enhanced persistent emission due to the Poynting–Robertson drag(the $f_a$-model)or the reflection emission illuminated by the burst(the reflection model). These two models provided similar burst parameters and are therefore difficult to distinguish.

To calculate the local accretion rate, we use the following equation: 
\begin{equation}
\begin{split}
    \Dot{m}
    &=\frac{L_\mathrm{ {per}}(1+z)}{ 4\pi R_{\rm NS}^{2}(GM_{\rm NS}/R_{\rm NS})}\\
    &\approx 1.7\times 10^{3}\biggl(\frac{F_\mathrm{{per}}}{4\times10^{-10}\mathrm{~ergs ~ cm^{-2}~s^{-1}}}\biggr)\biggl(\frac{d}{8\mathrm{~ kpc}}\biggr)^{2}\\
    &\quad\times\ \biggl(\frac{M_{\rm NS}}{1.4M_{\odot}}\biggr)^{-1}\biggl(\frac{1+z}{1.31} \biggr)\biggl(\frac{R\mathrm{_{NS}}}{10\mathrm{~ km}}\biggr)^{-1}\mathrm{g~cm^{-2}}\mathrm{~s^{-1}},\label{eq:lo_accration}
\end{split}      
\end{equation}\\ 
where $F_{\rm per}$ is the persistent flux and $d$ is the source distance at 8 kpc. The mass and radius of NS are $M_\mathrm{NS}=1.4M_\odot$ and $R_\mathrm{NS}=10$ km, and therefore the gravitational redshift on the NS surface is, $z\approx0.31$. We obtained
 $\dot{m}=1.63\pm{0.01}\times10^3~\mathrm{g~cm^{-2}~s^{-1}}$ for a distance of $8$ kpc \citep{01Paltrinieri}. The local Eddington accretion rate is $\dot{m}_{\rm Edd}=(8.8\times10^{4})[1.7/(X+1)]~\mathrm{g~cm^{-2}~s^{-1}}$, where $X$ represents the accreted hydrogen fraction. Therefore, the long burst of the UCXB source 4U~1850-087 occurred at a mass accretion rate of $\dot{m}\sim 1\%~\dot{m}_{\rm Edd}$.

\subsection{Burst parameters}
\begin{figure}[ht!]
\includegraphics[width=\hsize]{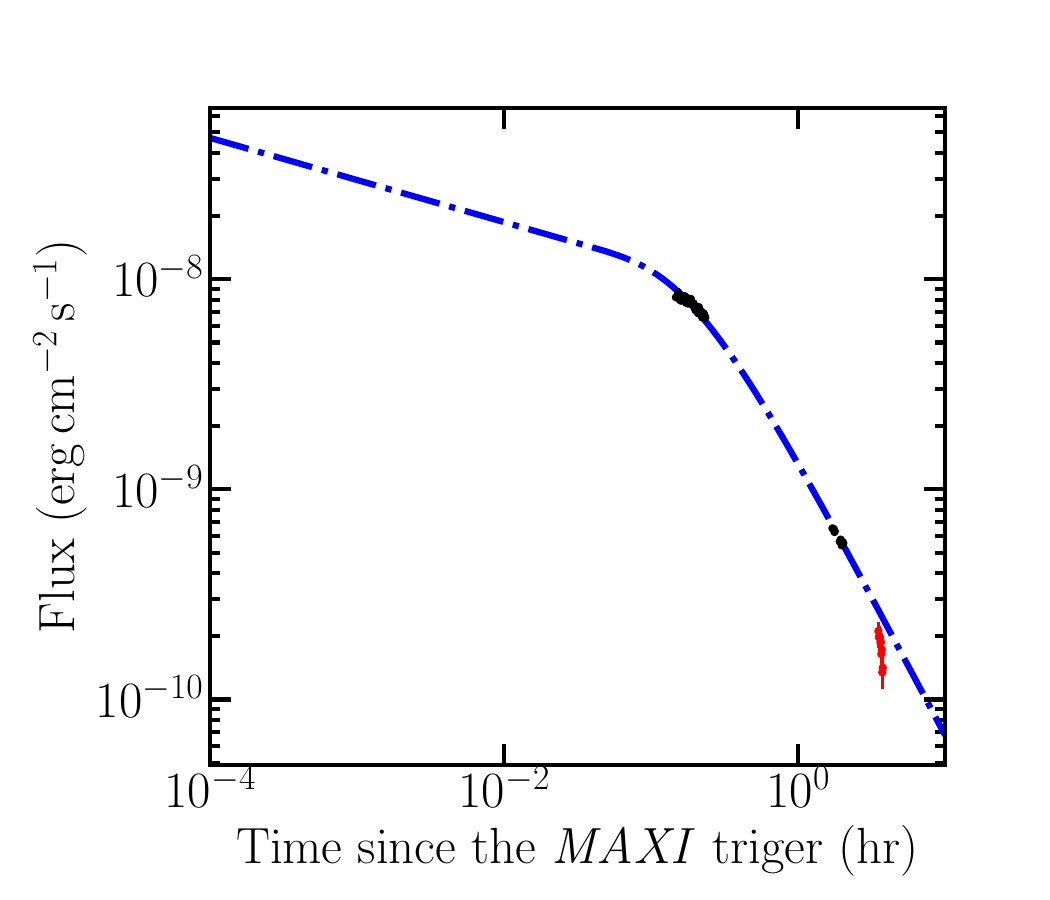}
\caption{The best-fitted burst decay flux by the model from \citet{Cumming04a}. The burst flux was estimated from the $f_a$-model. The black points represent the data in regions I and II. The red points represent the region III data excluded from the fit. We use the trigger time of MAXI as the burst peak time.} \label{fig:decay}
\end{figure}

The decay of the long X-ray burst flux can be well described with the analytic expression by \citet{Cumming04a} and \citet{Cumming06}, leading to constraints on the energy release per unit mass $E_{17}$ in units of $10^{17} \mathrm{~ergs~g^{-1}}$,  and the ignition column depth $y_{12}$ in units of $10^{12}\mathrm{~g~cm^{-2}}$. 
We used the trigger time from MAXI as the burst peak time about 0.15 hr earlier than the first NICER observation after the burst. We obtained a poor $\chi_{\nu}^2(\rm{dof})=2.94(47)$ by using all three regions. The burst fluxes in region III contribute the most significant residuals, which have also been shown in other long bursts \citep[see,e.g.,][]{Cumming04a,Cumming06}. Therefore, we only fit the fluxes from regions I and II. The fit is improved to an acceptable $\chi_{\nu}^2(\rm{dof})=1.02(37)$.
We show the $f_{\rm a}$-model bolometric flux with the best-fitted model from \citet{Cumming04a} and \citet{Cumming06} in Figure~\ref{fig:decay}. We estimated  $E_{17}=0.718\pm{0.001}$, $y_{12}=0.035\pm{0.001}$, and the burst peak flux $F_\mathrm{peak}=1.88\pm{0.07}\times10^{-8}\mathrm{~erg~cm^{-2}~s^{-1}}$.
The burst peak flux is less than the Eddington flux value, that is, $F_\mathrm{Edd}\approx3.8\times10^{-8}~\mathrm{erg~cm^{-2}~s^{-1}}$ in helium-rich environment \citep{Kuulkers03}.

We then use 
\begin{equation}f_\mathrm{{b}}=\frac{4\pi y_{\rm ign} R_{\rm NS}^{2}Q\mathrm{_{nuc}}}{{4\pi d^{2}(1+z)}}, \label{eq:ign}\end{equation} 
to calculate the burst fluence from the estimated ignition column depth $y_{\rm ign}$, where $Q\mathrm{_{nuc}}\approx 1.31 \mathrm{~MeV~nucleon^{-1}}$ for a given hydrogen fraction of $X=0$ \citep{Goodwin19}. 
We obtained the burst fluence $f_\mathrm{{\textit{b}}}= 5.3 \pm{0.2}\times10^{-5}\mathrm{~erg~cm^{-2}}$. We then calculated the burst decay time $\tau=f_{\rm \textit{b}}/F_{\rm peak} =0.78 \pm{0.04}$ hr, which is higher than the exponential decay time of 0.55 hr fitted from the NICER light curve without the first 0.15 hr data.  
All the burst parameters are reported in Table~\ref{table:burstpar}. 
For a comparison, in Table ~\ref{table:burstpar} we added two additional intermediate-duration bursts' parameters observed by MAXI and \swift\ \citep{Serino16,zand14}. The burst fluence and total released energy are rescaled to a distance of 8 kpc.  Our studied burst compared to the other two bursts has a smaller peak flux, a similar total burst emitted energy, and a longer burst decay time. The predicted recurrence time is obtained via the relation $\Delta t_{\rm rec}=y(1+z)/\dot{m}$. We found that the observed recurrence time, $\Delta T_{\rm rec}$, of the long X-ray burst observed in 2015 is higher than the predicted value. The discrepancy may be caused by the poorly calibrated flux from maxi, i.e., the burst peak flux is two times higher than the Eddington limit. 
The predicted recurrence time of the long X-ray in our work, 0.87 yr, is lower than the observed value, 7.76 yr, which means that some long or normal X-ray bursts were likely missed during 2015-2023 due to poor coverage of the observations. 

\begin{table*}
\begin{center} 
\caption{The  parameters of long X-Ray bursts from 4U 1850--087. \label{table:burstpar}}
\begin{tabular}{ccccccc} 
\hline
\hline 
Burst Date&
$F_{\rm peak}$&
$E_{\rm \textit{b}}$&
$f_{\rm \textit{b}}$&
$\tau$&
$\Delta t_\mathrm{rec}\equiv y(1+z)/\dot{m}$&$\Delta T_\mathrm{rec}^\mathrm{a}$\\
(DD-MM-YYYY)&$(10^{-8}~\mathrm{erg~cm^{-2}~s^{-1}})$&$(10^{41}~\mathrm{erg})$&$(10^{-5}~\mathrm{erg~cm^{-2}})$&(hr)&(yr)&(yr)\\\hline

10-03-2014$^\mathrm{b}$&10.7$\pm{0.9}$&$7.9\pm{0.7}~$&$10.3\pm{0.9}$&0.27&1.70&-\\
09-05-2015$^\mathrm{b}$&7.8$\pm{0.5}$&$14.8\pm{0.9}~$&$19.3\pm{1.2}$&0.71&3.18&1.16\\
11-11-2015$^\mathrm{c}$&-&-&-&-&-&0.51\\
15-08-2023&1.88$\pm{0.07}$&4.1$\pm{0.2}$&5.3$\pm{0.2}$&0.78$\pm{0.04}$&0.87$\pm{0.03}$&7.76\\
\hline\\
\end{tabular} 
\end{center}
$^\mathrm{a}$~ The calculated $\Delta T_\mathrm{rec}$ is the time interval between two consecutive observed bursts. However, $\Delta T_\mathrm{rec}$ represents the upper limit due to the data gap. 

$^\mathrm{b}$ The data were taken simply from \citet{zand14} and \citet{Serino16} and $E_{\rm \textit{b}}$ and $f_{\rm \textit{b}}$ are rescaled using  the distance of 8 kpc. 

$^\mathrm{c}$ The trigger timeS of bursts were taken from the MAXI novae web page.
\end{table*}

\subsection{Burst fuel}
The decay time of the burst from 4U~1850--087 on 2023 August 15, $\tau \sim$ 0.78 h,  belongs to the longest intermediate-duration bursts to the shortest superbursts. 
The burst decay times in the longest intermediate-duration bursts are, for example, 0.8 hr from SAX J1712.6--3739 \citep{zand19A&A} and 0.71 hr from 4U 1850-087 \citep[see Table~\ref{table:burstpar},][]{Serino16}. The burst durations of the shortest superbursts are, for example, 0.7 and 1.0 hr from GX 17+2, 1.2 hr from Ser X--1, and 2.1 hr from 4U 0614+091 \citep{Keek10}. The preburst mass accretion rates of GX 17+2 and Ser X--1 are $>20\%~\dot{m}_{\rm Edd}$, while the accretion rate of 4U 0614+091 is $<1\%~\dot{m}_{\rm Edd}$. 
The donor star of 4U~1850--08 is a helium white dwarf \citep{Deloye03,Falanga08}, which means that the accretion matter includes a large pile of helium. At a low accretion rate $\sim 1\%~\dot{m}_{\rm Edd}$, the carbon is difficult to accumulate \citep{Cumming06}.

The superburst from 4U 0614+091 occurred at the same low mass accretion rate as the burst observed from 4U 1850--087. However, the superburst decay time from 4U 0614+091 was about 2 hr compared with 0.78 hr for the burst in 4U 1850--087.  At the same accretion rate, we also found consistency between the observed and calculated recurrence times of the other two long bursts in 4U 1850--087 (see Table~\ref{table:burstpar}). Therefore, we propose that the studied burst from 4U 1850--087 is the longest intermediate-duration burst compared with other samples  \citep{zand14,Serino16}. 
We also calculated the ratio of the total persistent flux between two bursts to the total energy emitted from the burst $\alpha=(F_\mathrm{pers}\times \Delta T_\mathrm{rec})/f_\mathrm{\textit{b}}\sim117$, which is the shortest observed recurrence time of 0.51 hr used in Table~\ref{table:burstpar}. When we assume that all accreted fuel burned during the burst, we calculated the theoretical $\alpha=44~M_\mathrm{NS}R_\mathrm{NS}^{-1}(Q_\mathrm{nuc}/4.4\mathrm{MeV/nucleon})^{-1}\approx121$  \citep{Falanga08}. The calculated $\alpha$ from measurable quantities is close to a pure helium burst, supporting the burst as an intermediate-duration burst.

\subsection{The origin of the 1 keV emission line}
We found an emission feature near 1 keV in the burst spectrum, which can be best fitted by adding a Gaussian component. This line can only be detected in region I, but not in regions II and III. By using the tool \texttt{simfest}, we estimated the significance of the emission line, at $>7\sigma$, which proves that it is not caused by random fluctuations. Therefore, we propose that this line is caused by an astrophysical origin and not an instrumental effect.

Indeed, for the same source, 4U 1850-087, a similar weaker emission line of around 1 keV was also detected during an intermediate-duration burst \citep[see][]{Zand17}{}{}. Similar emission lines have also been detected in a few other NS LXMBs, e.g., in IGR J17062–6143, SAX J1808.4–3658, and 4U 1820–30, respectively \citep{Bult21, Bult19, Strohmayer19}. These emissions lines, from photoionized plasma of about 1 keV, are proposed to arise from the accretion disk illuminated by the burst blackbody emission. The involved processes include, for example, Fe L-shell transitions or Ly$\alpha$  transition of Ne X at 1.022 keV \citep{keek17,Degenaar_2013, Bult21} .
We note that the emission line around 1 keV was detected in 4U~1820--30 during the PRE phase, which was produced from the burst-driven wind \citep{Strohmayer19}. Considering that we miss the photosphere expansion in time-resolved spectra, the 1 keV emission line is likely produced in the accretion disk.

\section{summary}\label{sec:sum}
In this work, we have studied the long-duration burst from 4U 1850--087 observed by NICER and MAXI. We found that part of the burst spectra deviate from the blackbody model. We improved the fit by considering the reflection of the burst emission from the accretion disk or the enhanced persistent emission due to the Poynting-Robertson drag. We also found the 1 keV emission line when the reflection component is significant. We propose that the emission line originates in the Fe L-shell transitions or Ly$\alpha$ transition of Ne X at 1.022 keV from the accretion disk. From the time-resolved burst spectroscopy during the burst decay, we obtained the burst decay time, $\tau \approx 0.78$ hr, the burst fluence, $E_\mathrm{\textit{b}} \approx 4.1 \times 10^{41}$ erg, and the ignition column depth, $y_{\rm ign}\approx 3.5 \times10^{10}~{\rm g~cm^{-2}}$. Therefore, we propose that the burst is an intermediate-duration burst powered by the unstable burning of pure helium in the deep layer.

\begin{acknowledgments}
We thank the referee for the valuable comments that improved our manuscript.  This work was supported by the Major Science and Technology Program of Xinjiang Uygur Autonomous Region (No. 2022A03013-3). Z.S.L. and Y.Y.P. were supported by National Natural Science Foundation of China (12273030, 12103042). This work made use of data from the High Energy Astrophysics Science Archive Research Center (HEASARC), provided by NASA’s Goddard Space Flight Center. 

\end{acknowledgments}

\vspace{5mm}

\bibliography{name}{}
\bibliographystyle{aasjournal}

\end{document}